\documentclass[preprint2]{aastex62}
\usepackage{lineno}
\usepackage[version=4]{mhchem}
\usepackage{breqn}
\usepackage{amsmath}

\newcommand{\kms}{$\rm km~s^{-1}$}

\graphicspath{{./}{figures/}}

\shorttitle{Orion source I Disk Structure}
\shortauthors{Wright et al.}
\begin{document}
\title{Structure of the Source I disk in Orion-KL}
%\title{A Binary Protostar in Orion ?}
%\title{Observations of Orion Source I Disk and Outflow Interface}
%\title{Orion Source I as the disk driven outflow paradigm}
%% The \author command is the same as before except it now takes an optional
%% arguement which is the 16 digit ORCID. The syntax is:
%% \author[xxxx-xxxx-xxxx-xxxx]{Author Name}
%%
%% This will hyperlink the author name to the author's ORCID page. Note that
%% during compilation, LaTeX will do some limited checking of the format of
%% the ID to make sure it is valid.
%%
%% Use \affiliation for affiliation information. The old \affil is now aliased
%% to \affiliation. AASTeX v6.2 will automatically index these in the header.
%% When a duplicate is found its index will be the same as its previous entry.
%%

%% Use \email to set provide email addresses. Each \email will appear on its
%% own line so you can put multiple email address in one \email call. A new
%% \correspondingauthor command is available in V6.2 to identify the
%% corresponding author of the manuscript. It is the author's responsibility
%% to make sure this name is also in the author list.
%%
%% While authors can be grouped inside the same \author and \affiliation
%% commands it is better to have a single author for each. This allows for
%% one to exploit all the new benefits and should make book-keeping easier.
%%
%% If done correctly the peer review system will be able to
%% automatically put the author and affiliation information from the manuscript
%% and save the corresponding author the trouble of entering it by hand.

\correspondingauthor{Melvyn Wright}
\email{melvyn@berkeley.edu}

\author{Melvyn Wright}
\affiliation{Department of Astronomy, University of California, 501 Campbell Hall, Berkeley CA 94720-3441, USA}
\author{John Bally}
\affil{CASA, University of Colorado, 389-UCB, Boulder, CO 80309, USA}
\author{Tomoya Hirota}
\affiliation{Mizusawa VLBI Observatory, National Astronomical Observatory of Japan, Osawa 2-21-1, Mitaka, Tokyo 181-8588, Japan}
%\author[0000-0001-6765-9609]{Richard Plambeck}
%\affiliation{Radio Astronomy Lab, University of California, 501 Campbell Hall, Berkeley CA 94720-3441, USA}
\author{Kyle Miller}
\affiliation{Department of Astronomy, University of California, 501 Campbell Hall, Berkeley CA 94720-3441, USA}
\author{Tyler Harding}
\affiliation{Department of Astronomy, University of California, 501 Campbell Hall, Berkeley CA 94720-3441, USA}
\author{Keira Colleluori}
\affiliation{Department of Astronomy, University of California, 501 Campbell Hall, Berkeley CA 94720-3441, USA}
\author{Adam Ginsburg}
\affil{Department of Astronomy, University of Florida
211 Bryant Space Science Center
P.O Box 112055, Gainesville, FL 32611-2055 USA}
\affil{National Radio Astronomy Observatory, Charlottesville, VA 22903, USA}
\author{Ciriaco Goddi}
\affil{Leiden Observatory, Leiden University, P.O. Box 9513, 2300 RA Leiden, The Netherlands}
\author{Brett McGuire}
\affil{Department of Chemistry, Massachusetts Institute of Technology, Cambridge, MA 02139, USA}
\affil{National Radio Astronomy Observatory, Charlottesville, VA 22903, USA}

%\author{etal.}

%% Mark off the abstract in the ``abstract'' environment. 
\begin{abstract}
This paper analyses images from 43 to 340~GHz to trace the structure of the Source I disk in Orion-KL with $\sim$ 12 AU resolution. The data reveal an almost edge-on disk with an outside diameter $\sim$ 100 AU which is heated from the inside. 
%Images of the spectral index distributions show an extensive region with spectral index $<$ 2 along the minor axis.
 The high opacity at 220 - 340 GHz hides the internal structure and presents a surface temperature $\sim$ 500 K. Images at 43, 86 and 99 GHz reveal structure within the disk. At 43~GHz there is bright compact emission with brightness temperature $\sim$ 1300 K. Another feature, most prominent at 99 GHz, is a warped ridge of emission.
The data can be explained by a simple model with a hot inner structure, seen through cooler material.
A wide angle outflow mapped in SiO emission  ablates material from the interior of the disk, and extends in a bipolar outflow over 1000 AU along the rotation axis of the disk. SiO $v=0$ $J=5-4$  emission appears to have a localized footprint in the warped ridge. 
These observations suggest that the ridge is the working surface of the disk, and heated by
accretion and the outflow. The disk structure may be evolving,
with multiple accretion and outflow events. We discuss two sources of variability: 1) variable accretion onto the disk as Source I travels through the filamentary debris from the BN-Source I encounter $\sim$ 550 yr ago; and 2) episodic accretion from the disk onto the protostar which may trigger multiple outflows.
The warped inner disk structure is direct evidence that SrcI could be a binary experiencing episodic accretion.

\end{abstract}

%% Keywords should appear after the \end{abstract} command. 
%% See the online documentation for the full list of available subject
%% keywords and the rules for their use.
\keywords{radio continuum: stars --- radio lines: stars --- stars: individual (Orion source I)}

%% We recommend that authors also use the natbib \citep
%% and \citet commands to identify citations.  The citations are
%% tied to the reference list via symbolic KEYs. The KEY corresponds
% * <plambeck@berkeley.edu> 2018-08-15T17:41:25.565Z:
%
% ^.
%% to the KEY in the \bibitem in the reference list below. 

\section{Introduction} \label{sec:intro}

The Kleinmann-Low Nebula in Orion, at a distance 415 pc   \citep{Menten2007,Kim2008,Kounkel2018}, is the nearest HII region % or nebula %interstellar cloud sounds kinda vague, molecular cloud could be alternative but there is a lot of ionized atomic stuff floating around %interstellar cloud
in which massive ($M >8$~$M_{\odot}$) stars are forming. The two most massive objects in this region, Source~I (SrcI) and the Becklin-Neugebauer Object (BN), appear to be recoiling from one another at 35-40~\kms\ \citep{Rodriguez2005, Gomez2008, Goddi2011}, suggesting that they were ejected from a multiple system via dynamical decay approximately 550 years ago \citep{Bally2020}. % should match abstract?
SrcI has an estimated mass of $\sim$15~$M_{\odot}$ \citep{Ginsburg2018}, with a rotating accretion disk, a hot inner core obscured by a dusty outer envelope, and a molecular outflow that is prominent in shock-tracing SiO in several rotational-vibration levels.
 
The region around SrcI has been well studied since the 1980s
\citep{Hirota2014, Plambeck2016}; it is an active source associated with variable SiO and \ce{H2O} masers \citep{Reid2007, Goddi2009, Plambeck2009, Matthews2010, Goddi2011,  Niederhofer2012, Greenhill2013}.

% SrcI and evolved stars have several similar properties.
% however the identification of SrcI as an evolved object is not compatible with 
%{\bf 
The SiO emission and other properties of SrcI are reminiscent of similar features often observed in evolved stars.
However, the density of AGB stars within $\sim$ 1 kpc of the sun is low, $\sim$ 25  kpc$^{-2}$ (e.g., \citet{Jura1989},
and the chance of one located at the center of the Orion cluster being captured and dynamically ejected in the BN-SrcI encounter is small.
Although SrcI shows SiO maser emission similar to evolved stars, SiO vibrationally excited masers have been detected in several other high-mass star-forming regions (\citep{Hasagawa1986}, \citep{Higuchi2015}, \citep{Ginsburg2015}, \citep{Cho2016}, \citep{Cordiner2016}, \citep{Kalenskii2010}).
The presence of a disk-outflow system \citep{Hirota2017} indicates that SrcI is accreting, confirming its nature as a young, forming star.
%}

% The volume density of AGB stars in our part of the Galaxy is low, ~25/sq kpc (e.g., https://ui.adsabs.harvard.edu/abs/1989ApJ...341..359J/abstract), or ~0.25\%
% chance of one being in the Orion cloud region (assuming area 10x10 pc).  That's just areal projection; if we consider volume density, it's even lower likelihood.  Then, the chance of being located at the center of a cluster (see Justin Otter's paper), or being captured and dynamically directed there, is miniscule.

% The presence of NaCl and KCl should not be interpreted as evidence for an evolved state.  There are several other high-mass YSOs with detected NaCl and KCl now, so these lines are not unique to evolved stars.  Furthermore, no evolved star has detected high-vibrationally-excited states of NaCl and KCl, while SrcI does - this difference suggests that SrcI is not of the same class.  And further still, the morphology of the NaCl and KCl emission around SrcI is clearly disk-like, while in AGB stars, as far as I know, it's part of the spherical(ish?) wind.

% Similarly, while SiO masers in HMSFRs are rare, there are now ~10 known

% A brief historical review is called for.

% The SiO emission from source I suggested that it might be associated with an evolved star.
% ii) SiO outflows have now been identified with YSO
% iii) the BN- source I encounter, and stripped disk around source I is consistent with a YSO
%  The nature of a compact binary is not clear. The dynamical modeling favors two 10 M0 Objects

In this paper, we analyse high resolution continuum images from 43 to 340~GHz to trace the structure of the Source I disk at $\sim$ 30 mas (12 AU) resolution.
%show bright, high velocity SiO J=5-4 line
%reveal %a nonaxisymmetric structure in the disk linked to high energy molecular lines previously used to trace magnetic polarization and the rotating
 %emission extending in a 
 %bipolar outflow originating in a disk wind %rooted 
 %at the centrifugal radius $\sim$ 20-40 au.
 %, between cusps of low brightness continuum in the 99~GHz emission at the outer edge of the disk.
 %\citet{Basri1997} demonstrated the possibility for periodic accretion events in an eccentric binary orbit appearing as prominent flares at periastron.
%Most notably, SiO J=5-4 and H$_2$O at 230 GHz have highly collimated outflows anchored in the structure of interest at small radii ($<$ 25 au). This suggests an accretion-related event %observable in the structure of the circumbinary disk and outflow from SrcI.
%that may be related to binary-disk interactions, instability with spiral arms enhancing accretion, or magnetic reconnection.
% an edge-on disk evidently accreting onto SrcI, traced by a twisted ridge, and ablated into a hollow cone structure. We also analyze images from \citet{Wright2020} to interpret an unexpectedly low spectral index along the minor axis of the disk as further evidence of a hot inner core within the optically thick dusty outer shell.

The paper is organized as follows:  \S2 presents the observations and data reduction, \S3 discusses the results for the SrcI disk, \S4 summarizes the conclusions.

\section{Observations and Data Reduction}

We used JVLA and ALMA data at 43, 86, 99, 223 and 340 GHz to make 30 mas resolution images of the continuum emission from Orion Source I (SrcI). 
Table~1 provides a summary of the observations, including project codes and synthesized beam FWHM with robust= {-2} weighting of the uv-data.The \textsc{Miriad} software package \citep{Sault1995} was used to image and analyze all the data in this paper.

We present a new continuum image at 86 GHz using the spectral line data described in \citet{Hirota2020}. 
The data were calibrated using observatory supplied scripts. 
We used averaged spectra of the uv-data to identify line-free channels to image the SrcI disk in continuum emission.
% We averaged the spectra over channels with no spectral line emission in order to make continuum images.
% copy pdf plots to ~/public_html/UG_Research_2020/UG_research
%  30jan2021 make channel images for 86 GHz spectral windows. imaging all channels would take a long time, so we average the spectra over channels with no spectral line emission in order to make continuum images. the line emission is in channels  97 to 154 (approximately !). compare spw0 with 16-bit uvdata, with spw0a with 32-bit uvdata. imlist options=stat to see total flux, peak and rms and which image planes have line emission. plot to see image planes with and without line emission. average the off-line image planes and subtract to get line-continuum and RMS estimate
%The data include 4 spectral windows, each with 59 MHz bandwidth and 480 channels, and 5 spectral windows, each with 59 MHz bandwidth and 240 channels.
%{\bf We imaged each of these spectral windows.}
The data include 4 spectral windows, each with 59 MHz bandwidth and 480 channels, and 5 spectral windows, each with 59 MHz bandwidth and 240 channels.

%{\bf
We averaged the line-free channels on both ends of each
spectral window resulting in 18 averages of line-free uv-data with frequencies and bandwidths given in Table 3.
%spectral window  frequency  bandwidth
% 0  86.263704   86.223543  18.3
% 1  86.867220   86.827058  18.3
% 8  86.263376   86.223215  18.3
% 9  86.866891   86.826730  18.3
% 2  84.763768   84.728856  23.4
% 4  85.776890   85.741978  23.4
% 5 85.184697   85.149785  23.4
% 10 84.763444   84.728532  23.4
% 12 85.776567   85.741655  23.4

 We imaged these 18 averages and plotted the total flux and peak from Gaussian fits versus frequency in order to verify the consistency of the continuum images.
  After subtracting the average continuum image, the residual images were consistent with Gaussian noise distributions. Thus, the 86 GHz
continuum image is not significantly contaminated by unidentified spectral line emission.
%}

% # spectral index of total flux c/f with 99 (58 +/- 6 mJy) and 43 GHz (10 +/-1 mJy) wright2020
% calc '(86/99)**2*58' = 43.76778
% calc '(86/43)**2*10'  = 40.00000

% 86GHz_Flux.csh
% /alma_scr/wright/OriPol2018
% # Gaussian fit to 86 GHz images
% foreach SPW ( 0a 0 1 8 9 2 4 5 10 12 )
%  imfit  in=spw$SPW.I.cm.ave object=gauss 'region=arcsec,box(0.2,-0.2,-0.1,0.1)'
% end

% # plot flux and peak versus frequency using wip

% wip 86GHz_Flux.wip -d 86GHz_Flux.ps/cps -x
% ps2pdf 86GHz_Flux.ps
% cp 86GHz_Flux.pdf ~/public_html/UG_Research_2020/UG_research/

We made images with several different weightings of the uv-data; using robust= {-2} weighting to produce a more uniform weighting of the uv-data to emphasize compact structures, and robust= {+2} weighting for better sensitivity to low brightness, more extended emission.
% copy the results to ~/public_html/UG_Research_2020/UG_research. plot the continuum averages. clean and plot the continuum averages. copy the results to ~/public_html/UG_Research_2020/UG_research. 31jan2021 - plot whole image. Clean does not include BN ~ 8 arcsec NW from source I. Need to make imsize=2048 options=double to include BN. Sidelobes from BN could contaminate source I, although the full field looks clean. An image including BN would be good for flux check on source I, and for the BN paper.

\begin{figure}
% trim left bottom right top
\includegraphics[width=1.0\columnwidth, clip, trim=3cm 3.7cm 2.5cm 1cm]{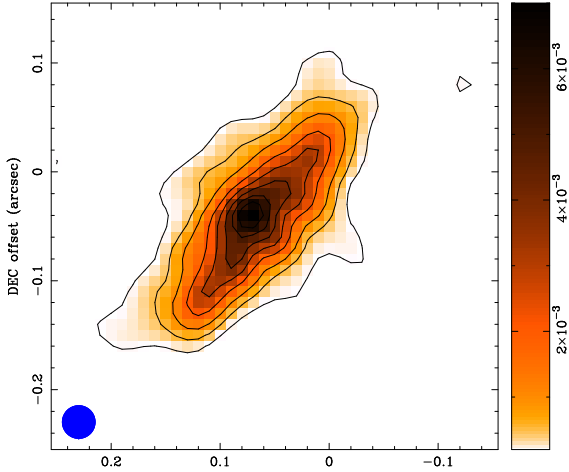}
\caption{Source I continuum emission at 86~GHz.
 Contour levels: 0.2 0.5 1 2 3 4 5 6 7 mJy/beam.
The 30 mas convolving beam FWHM is indicated in blue in the lower left.
\label{fig:86GHz_image}}
\end{figure}

% mchw 01feb2021. Average off-line channels, then make mfs image from averaged uv-data. spw with 480 channels. Average in 2 groups of 150 channels. the line emission is in channels 193 to 288. spw with 240 channels. Average in 2 groups of 96 channels. the line emission is in channels  97 to 154
We then made a Multi-frequency Synthesis (MFS) image from the averaged line-free uv-data. % with robust= {-2} weighting. 
The image includes the BN source $\sim$ 10 arcsec NW from  SrcI.
We deconvolved the image using the CLEAN algorithm over the full field of the image, including the BN source.
Sidelobes from BN could contaminate SrcI, although the full field looks clean. The image including BN also provides a flux check on SrcI, and will presented in another paper.
The MFS image of  SrcI at 86 GHz is shown in Figure~\ref{fig:86GHz_image}.

% mchw 13 march 2021 - revised figure
%/alma_scr/wright/alma2018/340GHz
%morph>grep B7-B6-B3_cm03.ps *.csh
%B7-B6-B3.csh:
% make 99 GHz image in K units
%cgdisp in=X438c.cont.memcm.03,B6.ms0.mfs.mem2cm.03.regrid.B7,cont.mfs.cm.03.regrid.B7.K labtyp=arcsec type=c,c,p options=full,noepoch,beambr,blacklab,wedge 'region=arcsec,box(.25,-.25,-.1,.1)' lines=2,2 cols1=2 cols2=4 device=B7-B6-B3_cm03.ps/cps levs1=50,100,200,300,400,500 levs2=50,100,200,300,400,500 slev=a,8.4925436E-05,a,3.8925558E-05 range=1.4e-3,8e-3,lin,-8 range=100,1100,lin,-8
%ps2pdf B7-B6-B3_cm03.ps
%cp B7-B6-B3_cm03.pdf ~/public_html/UG_Research_2020/UG_research/

%\begin{figure}
% trim left bottom right top
%\includegraphics[width=1.0\columnwidth, clip, trim=3cm 3.7cm 2.5cm 1cm]{figures/B7-B6-B3_cm03.pdf}
%\caption{Source I continuum emission at 99, 220, and 340 GHz.
%Red  contour levels at 340~GHz: 50,100,200,300,400,500 K.
%Blue contour levels at 220~GHz: 50,100,200,300,400,500 K.
%The pixel image shows the continuum emission at 99 GHz in K.
%The 30 mas convolving beam FWHM is indicated in blue in the lower right.
%\label{fig:B7-B6-B3}}
%\end{figure}

\begin{figure}
% trim left bottom right top
\includegraphics[width=1.0\columnwidth, clip, trim=3cm 3.9cm 2.5cm 1cm]{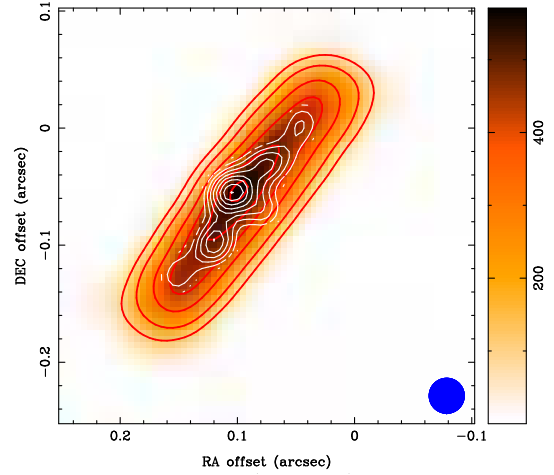}
\caption{Source I continuum emission at 99, 220, and 340 GHz.
White contour levels at 99~GHz: 50, 100, 200, 300, 400, 500, 600, 700, 800, 900, 1000 K. Red contour levels at 220GHz: 50, 100, 200, 300, 400, 500 K. The pixel image shows the continuum emission at 340 GHz in K. The 30 mas convolving beam FWHM is indicated in blue in the lower right.
\label{fig:B7-B6-B3}}
\end{figure}

% 
%start:
%# 26 may 2021
% /alma_scr/wright/VLA_pol/map.csh
%# white contours and thicker lines
%cgdisp in=B2.cm.03.imcat.ave.regrid.B3.adjust,86GHz.mfs.I.cm.03.regrid.B2,86GHz-43GHz.SI.cm.03 labtyp=arcsec 'region=arcsec,box(.2,-.2,-.1,.1)' options=full,blacklab,beambr,noepoch,3pix,wedge type=c,c,p slev=a,1.3668522E-06,a,5.4673219E-06 levs1=50,100,200,400,600,800,1000,1200,1400 levs2=50,100,200,400,600,800,1000,1200,1400  range=1,3,lin,-8 device=86GHz-43GHz_SI_cm_03.ps/cps cols1=1 cols2=4 lines=4,4,4,4,4,4
%ps2pdf 86GHz-43GHz_SI_cm_03.ps
%cp 86GHz-43GHz_SI_cm_03.pdf ~/public_html/UG_Research_2020/UG_research

\begin{figure}
% trim left bottom right top
\includegraphics[width=1.0\columnwidth, clip, trim=3cm 4.5cm 2.5cm 1cm]{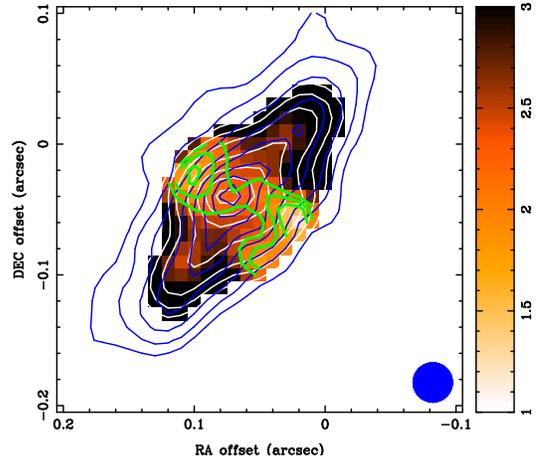}
\caption{Spectral Index 43~GHz to 86~GHz.
White contour levels at 43~GHz: 50, 100, 200, 400, 600, 800, 1000, 1200 K. Blue contour levels at 86~GHz: 50, 100, 200, 400, 600, 800, 1000, 1200 K. The color image shows the spectral index distribution. The thick green contours at SI=1.7, 2 and 2.3 delineate the region of lower SI along the minor axis
of the disk. The 30 mas convolving beam FWHM is indicated in blue in the lower right.
\label{fig:86Hz-43GHz_SI}}
\end{figure}

{\bf \begin{figure}
% trim left bottom right top
\includegraphics[width=1.0\columnwidth, clip, trim=3cm 4.5cm 2.5cm 1cm]{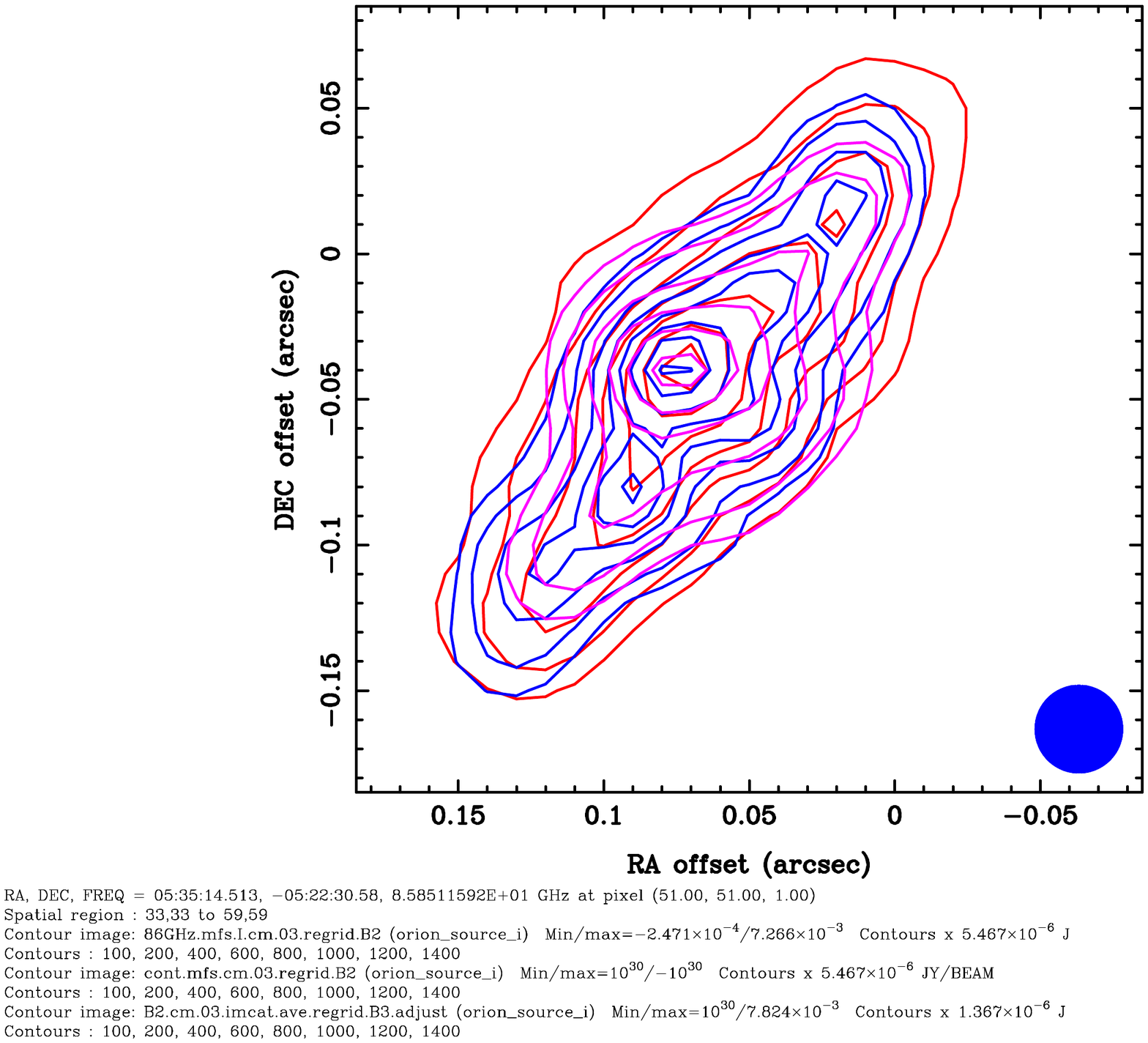}
\caption{43~GHz, 86~GHz, and 99~GHz structure.
Purple contour levels at 43~GHz: 100, 200, 400, 600, 800, 1000, 1200 K. Red contour levels at 86~GHz: 100, 200, 400, 600, 800, 1000, 1200 K. Blue contour levels at 99~GHz: 100, 200, 400, 600, 800, 1000, 1200 K. The 30 mas convolving beam FWHM is indicated in blue in the lower right.
\label{fig:43-86-99GHz}}
\end{figure}
}

% /alma_scr/wright/alma2018/340GHz
% morph>vi B6-B3.SI.csh
%imlist options=stat in=86GHz.mfs.I.cm.03.regrid.AlO
%calc '182904.6250*7.2884E-03' = 1333.082 K max
%cgdisp  labtyp=arcsec in=86GHz.mfs.I.cm.03.regrid.AlO,86GHz.mfs.I.cm.03.regrid.AlO,AlO_N=6-5.B6.spw0.mom0 options=full,beambr,noepoch,blacklab,wedge,relax type=p,c,c range=2e-4,7e-3,lin,-8 cols1=2 cols2=4 lines=2,2,2,2 levs1=.2,.5,1,2,3,4,5,6,7 slev=a,1e-3,a,6.4595311E-05 levs2=500,1000,1500,2000,2500,3000,3500,4000,4500 device=86GHz+AlO_N=6-5_B6.ps/cps

\begin{figure}
% trim left bottom right top
\includegraphics[width=1.0\columnwidth, clip, trim=3cm 4.4cm 3.2cm 1cm]{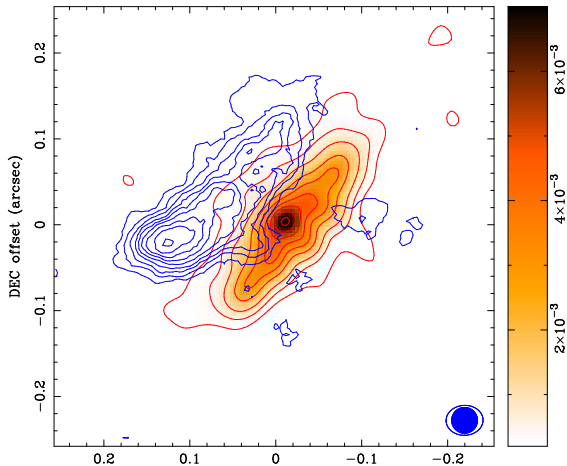}
\caption{AlO N=6-5 emission at 229.6938725 GHz
% with 2 sigma cut.
 overlayed on 86 GHz continuum emission. 
 Red contours show the 86 GHz continuum emission at 37 to 1280 K with a contour interval of 183 K. 
 %37K, 91K, 183K, ... 1280 K. contour interval 183 K.
Blue contours show the AlO $N=6-5$ emission integrated over -50 to 50~\kms. Lowest contour and contour interval 500 K km s$^{-1}$.  The 30 mas convolving beam FWHM for the 86 GHz continuum image is indicated in blue filled circle. The FWHM beam for the AlO N=6-5 is 0.04 $\times$ 0.03 arcsec in PA -88$^{\circ}$, are shown as the open ellipse in the lower right.
\label{fig:86GHz+AlO}}
\end{figure}

% morph>vi 86GHz.mfs+sio54.csh
% pwd /alma_scr/wright/OriPol2018
% imlist options=stat in=86GHz.mfs.I.cm.03.regrid_sio54
%  bunit: JY/BEAM  frequency:  86.243439    K/Jy:     182904.6250
  % cgdisp in=86GHz.mfs.I.cm.03.regrid_sio54,86GHz.mfs.I.cm.03.regrid_sio54,spw3.sio.cm-cont region=arcsec,kms,box'(.25,-.25,-.15,.15)(-20,-10),box(.25,-.25,-.15,.15)(26,36)' levs1=.2,.5,1,2,3,4,5,6,7 slev=a,1e-3 range=2e-4,7e-3,log,8 type=p,c labtyp=hms,dms options=wedge,blacklab,noepoch,full,beambl,3val csize=.7,.7,.7,.7 cols1=1 labtyp=arcsec device=86GHz.mfs.I.cm.03+sio54.ps/cps

%\begin{figure*}
% trim left bottom right top
%\includegraphics[width=2.0\columnwidth, clip, trim=3cm 4.3cm 2.7cm 1cm]{figures/86GHz_mfs_I_cm_03+sio54.pdf}
%\caption{SiO $v=0$ $J=5-4$  emission at 217.10498 GHz overlaid on 86 GHz continuum.
%Red contours show the high velocity SiO $v=0$, $J=5-4$ emission at 2 km s$^{-1}$ intervals. Contour interval 10 mJy/beam. (143 K x 2km/s). Beam FWHM 54 x 34 mas in PA 65 \degr.
%Blue contours show the 86 GHz continuum image. Contours at .2,.5,1,2,3,4,5,6,7  mJy/beam in a 30 mas FWHM beam. (37 91 183 366 549 732 915 1097 1280 K). 
 %\label{fig:86GHz+SiO54}}
%\end{figure*}

\begin{figure*}
% trim left bottom right top
\includegraphics[width=2.0\columnwidth, clip, trim=.5cm 4.3cm 0.5cm 1cm]{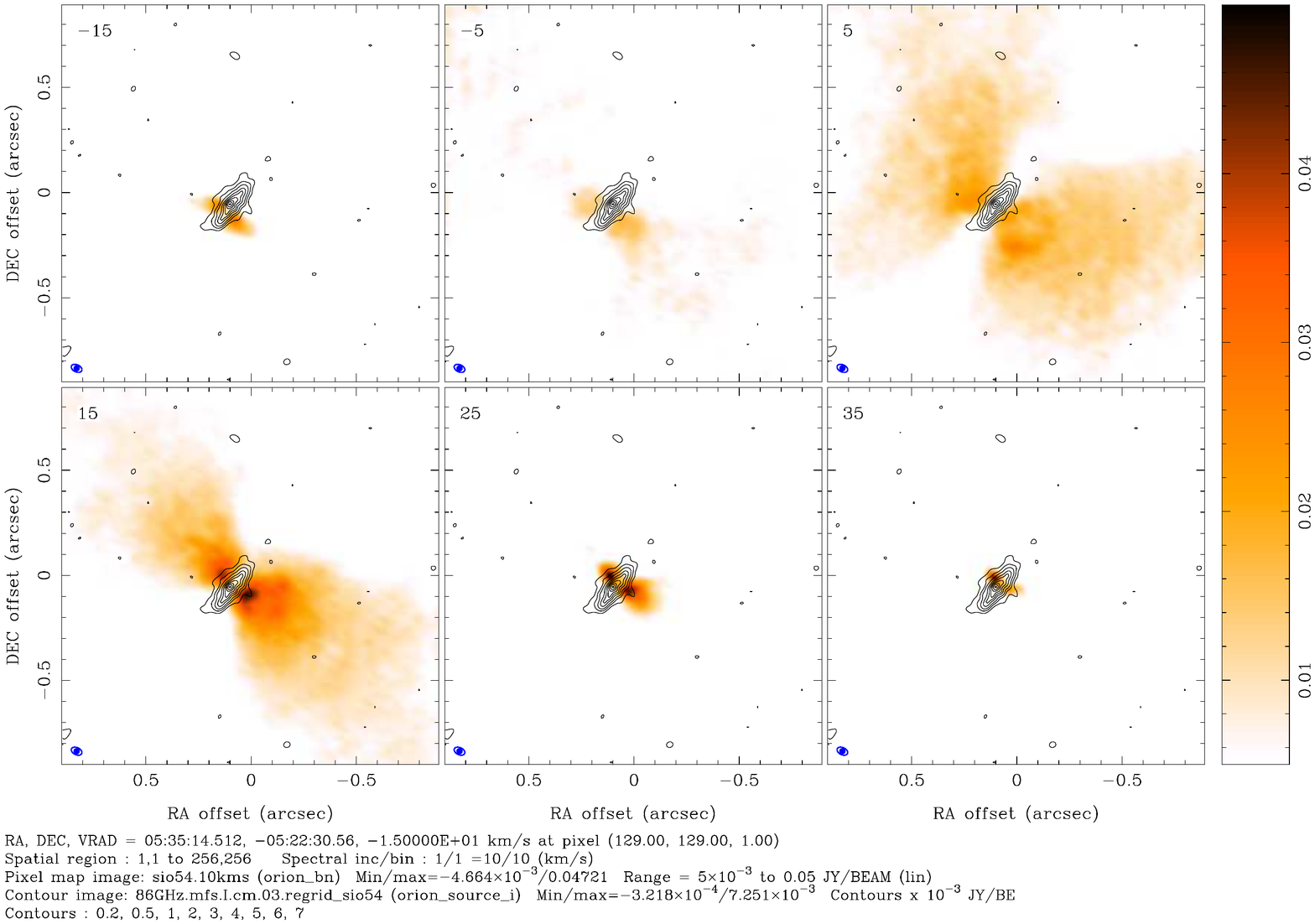}
\caption{SiO $v=0$ $J=5-4$  emission at 217.10498 GHz overlaid on 86 GHz continuum. The
color image shows the high velocity SiO $v=0$, $J=5-4$ emission at 10 km s$^{-1}$ intervals. Beam FWHM 54 x 34 mas in PA 65 \degr.
Red contours show the 86 GHz continuum image. Contours at .2,.5,1,2,3,4,5,6,7  mJy/beam in a 30 mas FWHM beam. (37 91 183 366 549 732 915 1097 1280 K). 
 \label{fig:86GHz+SiO54}}
\end{figure*}

%{\color{red} This section can be brief, refering to Hirota2020, Wright2020 and Ginsburg2018 for details.  Table~1 updated MCHW}

%{\color{red} description of SiO lines used belongs in section{THE SRC I OUTFLOW}}

\subsection{43 GHz and 99 GHz}
The 43~GHz and 99~GHz (ALMA Band 3; B3) observations and calibration are described in \citet{Hirota2020} and \citet{Wright2020}.
The 43~GHz JVLA observations included 16 wideband spectral windows each covering 116~MHz bandwidth. The wide band average with a mean frequency of 42.65~GHz and a bandwidth of 1.4~GHz excludes 4 spectral windows containing spectral line emission.% We made images with robust=-2 weighting of the uv-data, and deconvolved to 30~mas for comparison with the other images.

%The JVLA wideband setup is 64 channels covering 128 MHz; the reduction script trimmed off end channels because of analog filter rolloff. 
%The data were self-calibrated using a strong SiO $v=1$, $J=1-0$ maser feature at -3.5~\kms\  that was observed simultaneously in a narrowband window with spectral resolution 0.3 km s$^{-1}$.
%/alma_scr/wright/VLA_pol/map.csh 
% 11jan2019 remake with same imsize=3200, so can combine
% foreach SPW (0 1 2 3 4 5 6 7 8 9 10 11 12 13 14 15 )
% gpcopy vis=line.spw1.uv2 out=spw$SPW.uv

% invert vis=spw$SPW.uv map=spw$SPW.mp beam=spw$SPW.bm robust=0.5 options=mfs imsize=3200 cell=0.01
% SELFCAL  interval=0.1  vis=line.spw1.uv2 line=velocity,1,-3.5,.5,.5

% grep "#"  map.csh
% use VLA as EVLA not recognised by Miriad's telepar for PB etc.
% quoted SEFD= 1260 at 45 GHz. 
% copy 6s selfcal gains from sio v=1 maser into wide windows
% Note imcat channel order is 1.... then 2,3,...9

%\subsection{99 GHz}
ALMA observations at 99~GHz included 4 spectral windows, each with a bandwidth of 1.875~GHz and 960 spectral channels. 
The wide band average with a mean frequency of 99.275~GHz and a bandwidth of 7.5~GHz excludes channels with spectral line emission. 

%alma_scr/wright/OriPol2018/cont/cont.csh
% GPCOPY:   vis=spw0.uv   out=spw6.avg.uv  mode=copy options=nopol,nopass
% INVERT:   vis=spw6.avg.uv,spw7.avg.uv,spw14.avg.uv,spw15.avg.uv  line=chan,1,1,1 map=cont.mfs.mp beam=cont.mfs.bm stokes=I robust=-2 cell=.010 imsize=4096options=nocal,mfs
% SELFCAL:   offset=0.137,-0.065   refant=24
% Total number of solution intervals: 1509
% Rms of the gain phases (degrees):  28.1

\subsection{223 and 340 GHz}

The 223~GHz (ALMA Band 6; B6) and 340~GHz (ALMA Band 7; B7) observations and calibration are described in \citet{Ginsburg2018}.
Images of SrcI with 30 mas resolution at 99, 220, and 340 GHz \citep{Wright2020} are shown in figure~\ref{fig:B7-B6-B3}.
%These data were not self-calibrated. The { observational parameters} are given in Table 1.
% Justification for 30 mas: greater than lambda/2Dmax
% 300./43/36600e3*2e5/2 = 0.019 arcsec
% 300./86/16200e3*2e5/2  = 0.02
% 300./224/10500e3*2e5/2 = 0.012
% 300./340/10500e3*2e5/2 = 0.008

\subsection{Spectral Index Distribution}
%% TODO 
% use 43 and 86 SiO masers to trace density and temperature (extend model from above)
% elaborate on pumping mechanism (plasma?)
% mention potential period-luminosity relation; this could be a check for L, M and R

We made spectral index (SI) images between 86~GHz and 43~GHz images clipped at 25~K.
Figure~\ref{fig:86Hz-43GHz_SI} shows the spectral index computed from the ratio of continuum brightness at 86 and 43 GHz.
Spectral index images between 43, 99, 222, and 340 GHz are shown in \citet{Wright2020}.
%The RMS noise levels on the continuum images are 4~K, 3~K, 7~K, and 6~K, respectively; the images were clipped at 5\% of the peak for the spectral index calculation.  The figures show that spectral index $\sim$2 along the midplane of the disk, consistent with optically thick emission, almost certainly from dust \citep{Plambeck2016}.  The spectral index steepens to $\sim$3 at the edges and ends of the disk, indicating emission from optically thin dust.

In Figure~\ref{fig:86Hz-43GHz_SI}, we see that the central source is more prominent at 43~GHz, and also note the steepening of the spectral index at the ends of the disk between 43 and 86~GHz.

% Details:

% /alma_scr/wright/alma2018/340GHz
% B7-B3.SI.csh
% B6-B3.SI.csh

% scaling factor and estimate average SI from imdiff between B6 and B3 images
% calc 'log(3.5)/log(223/99.276)' 
% calc 'log(3.1)/log(223/99.276)'

We checked the individual spectral
windows at 43 and 86 GHz for any residual spectral line contamination in the averaged continuum images. After subtracting the average continuum images shown in Figure~\ref{fig:86Hz-43GHz_SI}, the individual spectral
windows are consistent with Gaussian noise distributions. Thus, the
continuum and spectral index images are not significantly contaminated by unidentified spectral line emission.
%At 43~GHz the RMS in the 12 wideband spectral windows is 14 K with maximum and minimum +/- 45 K.
%At 99~GHz the RMS in the four wideband spectral windows is 35 K with maximum and minimum +/- 100 K.
%calc '5e-5*280201.2812'
%# RMS =  14.01 K Max/Min +/- 1.5E-4 = 45 K
%cp B2.cm.imcat.subtract.pdf ~/public_html/UG_Research_2020/UG_research
%cp map2.csh ~/public_html/UG_Research_2020/UG_research/

Errors in the spectral indices are also
limited by the absolute flux calibration, resulting in an accuracy of $\sim$10\% for each frequency band.
%The errors in the spectral index images are limited by calibration errors and the flux density scale for each frequency.
% We estimated the absolute errors in the spectral index images from flux scale errors 10\% on each image, and 20\% in the flux ratio.
%A multiplicative error in the flux ratio is an additive error in the spectral index of +0.15 and -0.18 in the 340/99 GHz spectral index image, +0.22 and -0.28 in the 223/99 GHz image, and +0.11 and -0.13 in the 99/43 GHz image.
A least squares fit from 43 to 340~Ghz to the spectral index at the central position gives $1.6\pm.1$ %+/- 0.1,
whereas at the ends of the disk  the fitted spectral index is  $3.4\pm.3$. % +/- 0.3.

%The spectral index variations across the images are significant.

%{\color{red} We won't use these low resolution observations}
%In order to compare the SrcI outflow with the large scale structure associated with the SrcI/BN explosion mapped in CO by \citet{Bally2017}, we imaged other spectral lines that fell within the passband of these lower resolution observations (ADS/JAO.ALMA\#2013.1.00546.S).  The data were calibrated using the observatory-supplied CASA scripts. We made mosaic images from these data in a 30$''$ field around SrcI in  SO, SO$_2$, and SiO emission with 1.5 $\times$ 0.9 $''$ angular and 2~\kms\ velocity resolution. We used only the ALMA 12m array data in these images to filter out the large scale structure and enhance the filamentary structures.

\section{DISCUSSION}
%\subsection{THE SRC I DISK}
%{\color{red} brief description referring to these details in Wright2020. New FIGURE showing the non-thermal emission at 43 GHz, and the twisted ridge at 99 GHz.}

\subsection{Disk Structure}
   
\citet{Ginsburg2018} fitted the observed structure of the disk from B3 and B6 ALMA continuum observations at 50 and 20 mas resolution, respectively. They determined that the disk has a length of $\sim$100 AU, and  vertical FWHM height of $\sim$20 AU.  They also detected a compact source near the center of the disk, smeared parallel to the disk major axis.  

%The model residuals shown by \citet{Ginsburg2018} have a halo of emission at the $\sim$30 K level in the B6 image that may be from optically thin dust, as it was not seen in the B3 image. This halo of emission is evident in Figure~\ref{fig:B6-B3_SI}, where both B3 and B6 are plotted at the same logarithmic contour levels. 

Table~2 summarizes the results of Gaussian fits to the disk size from the 43 to 340~GHz images.  The disk major axis increases with observing frequency, which is expected as the dust optical depth increases.  The minor axis is largest at 43 GHz, however, indicating that the compact source is more prominent at lower frequencies.

%{\color{red}

Figure~\ref{fig:86GHz_image}, ~\ref{fig:B7-B6-B3}, ~\ref{fig:86Hz-43GHz_SI}, and ~\ref{fig:43-86-99GHz} show images of the source structure at 30 mas resolution visible from 43 to 340 GHz. Some sub-structures are evident:
1) A compact source most prominent at 43 GHz;
2) A spectral index, SI $\sim$ 3 at the ends of the disk along the major axis;
3) An extended region with spectral index, SI $<$ 2 along minor axis, and extensions along the major axis ridge; and, 
4) A warped ridge of emission at  43, 86, and 99 GHz---most prominent at 86 and 99 GHz, but still visible at 220 and 340 GHz.

We discuss these structures in the following section.

{\color{red}}

\subsubsection{Compact Source}

The peak brightness at 220 and 340 GHz coincides
with the compact source seen in the 43 - 99 GHz images. As argued in \citep{Ginsburg2018}, the
position is offset from the center of the disk,
and the source is extended, suggesting that it is a hot region in the inner disk, and not the location of the central source powering  SrcI. 
%{\bf The offset location and extended emission may correspond to the launching radius, 21 - 47 AU of the SiO outflow \citep{Hirota2017} seen in projection in the nearly edge on disk.}

\subsection{Spectral Index}

The spectral index images in \citep{Wright2020} also show
a spectral index (SI) $\sim$ 3 at the ends of the disk associated with dust emission, and the
region with SI $<$ 2 along the minor axis of the disk. Our new image at 86 GHz shows
an extended region with SI $<$ 2 along the minor axis, as well as a narrow region tracing the warped ridge which is evident in the 99 GHz and 86 GHz continuum images. We discuss three mechanisms which could produce an SI $<$ 2.

{\bf \subsubsection {a) Free-free emission and ionized jets.}}
\citet{Plambeck2016} analysed the continuum flux densities measured for SrcI from 4 to 690 GHz.
Above 43 GHz the flux density is consistent with emission from optically thick dust. At lower frequencies, the flux densities are higher than the $\nu^2$ fit, and free–free emission may become important where the dust emission becomes optically thin.  (see \citet{Plambeck2016} - Figure 14).

The H26$\alpha$ (353.6 GHz) and H21$\alpha$ (662.4 GHz) hydrogen recombination lines were not detected. 
\citet{Plambeck2016} used a velocity range 5 $<$ VLSR $<$ 23 km s$^{-1}$  to establish upper limits on the intensities of the recombination lines. The blue-shifted portion of recombination lines would be confused by an unidentified emission line at 353.648 GHz, and a strong absorption by a transition of SO$_2$ at
662.4043 GHz (see \citet{Plambeck2016} Figure 6). The data were integrated in a 0.2 $\times$ 0.2 $''$ box centered on SrcI.

No recombination lines were detected in the SrcI disk with an upper limit 5\% of the total continuum flux. 
Electron–Neutral Free–Free Emission
\citep{Reid2007} may be responsible for the higher flux densities below 43 GHz.  An ionized jet would have a spectral index $<$ 2. Although no  well collimated jet has been detected in SrcI, the extensions to the NE along the minor axis at 43 GHz, and a smaller one at 86 GHz (see Figure~\ref{fig:cont_mfs_cm03+sio54_10kms}) have a spectral index $\sim$ 1. 
 Recent radio continuum surveys have detected a number of high-mass YSOs associated with radio jets \citep{Purser2016}. They are thought to be driven from closer to the central YSOs given their higher velocities of a few 100 km s$^{-1}$.
 Similar to these other high-mass YSOs, there could be a compact radio jet in SrcI emanating from the inner region.

In NGC7538 IRS1, for example, \citet{Sandell2020} find that the size of the core decreases as $\nu^{-0.92}$, with a spectral index $\alpha$ = {0.87}, (see \citet{Sandell2020} figure 2), which is consistent with an ionized jet\citep{Reynolds1986}.
High resolution observations at longer wavelengths are required to determine if SrcI harbors an ionized jet.

The data do not preclude a small contribution to the continuum from free–free emission, but this would not account for the extended regions we see with SI less than 2.

% synchrotron - usually close to star surface, not extended as seen. no polarization detected \citep{Hirota2020}

{\bf \subsubsection{b) Non-thermal emission}}
Non-thermal emission may contribute to the observed spectral index of the SrcI disk. Synchrotron emission is known to develop in the circumstellar disk jets of massive proto-stars {\citep{Araudo2021}}.
\citet{Sanna2019} detected a nonthermal jet at 8 GHz which is driven by a young massive star in the star-forming
region G035.02+0.35.

%We made use of the NSF’s Karl G. Jansky Very Large Array (VLA) to observe this region at C, Ku, and K−1bands with the A- and B-array configurations, and obtained sensitive radio continuum maps down to an rms of 10 μJy beam. These observations allow for a detailed spectral index analysis of the radio continuum emission in the region, which we interpret as a proto-stellar jet with a number of knots aligned with extended 4.5 μm emission. Two knots clearly emit nonthermal radiation and are found at similar distances, of approximately 10 000 au, at each side of the central young star, from which they expand at velocities of several 0.4 × 10−4 M⊙ yr−1 km s−1 and values in the range 0.7–1.3 mG. hundred km s
However, this mechanism is unlikely in SrcI because a well formed jet is not observed.
Synchrotron emission is associated with a spectral index less than 1, and might also be observed close to the surface of the star. but not in the extended circumstellar disk as observed in  SrcI. \citet{Hirota2020} did not find any polarization in the SrcI disk, which might also be expected for synchrotron emission.

%without invoking a separate emission mechanism,
{\bf \subsubsection{c) Dust opacity}}
A simple explanation
could be a temperature gradient in the disk which reveals hotter material in the inner disk at lower frequencies where the opacity in the disk is lower. 
A hot inner disk is also suggested by the observations of high excitation SiO transitions \citep{Ginsburg2019} - see Figure 5, \citep{Kim2019}, \citep{Hirota2017}, and H$_2$O emission \citep{Wright2020}.

% \subsubsection{The Hollow Cone Structure}

Figure \ref{fig:86GHz+AlO} shows contours of the AlO N=6-5 emission integrated over -50 to 50~\kms,
overlayed on the 86 GHz continuum emission. The AlO emission forms a cone on the NE side of the disk, with a wide angle outflow
extending from the warped ridge in the 86 GHz continuum emission from the  SrcI disk. The distributions of AlO emission shown here is consistent with that presented by
\citet{Tachibana2019} for AlO N=13-12 and N=17-16 emission lines at 497 and 650 GHz, who suggest that AlO is formed at the base of the outflow, and condenses further out into the outflow. Our observations {with $\sim 4\times$ higher angular resolution,} show that AlO emission extends from deep within the SrcI disk, and may be produced by grain destruction. % at the disk surface.
%{ peaks} downstream of the \ce{H2O}, and %and oxidized by O released by the dissociation of \ce{H2O}, or released as AlO further out in the outflow than the \ce{H2O} emission. AlO  mapped in the outflow close to the disk, suggest that refractory grain cores as well as the grain mantles are destroyed. 
\citet{Lenzuni1995} investigated the evaporation of dust grains in protostellar cores. Carbon grains are destroyed at temperatures $\sim$ 800 -- 1150 K. Silicate grains are evaporated at temperatures $\sim 1300$ K, and AlO at $\sim$ 1700 K.
%SiO and AlO may be released directly from the grains, or may be formed in the gas phase by the oxidation of Si and Al. 
The distribution of AlO offers strong support for ablation of the disk into a hollowed out cone, leaving a flared disk  of cooler material which obscures the interior, with high dust opacity at higher frequencies. The strongly asymmetric distribution of AlO emission suggests that the heating source is associated with the compact source seen at 43 GHz on the east side of the disk.

  The inner surface must be $\sim$ 1700 K to evaporate AlO. The outer surface temperature of the disk is observed to be $\sim$ 500 K at higher frequencies. Radiation from the structure of the interior of the SrcI disk is absorbed in the exterior part of the disk. 

%CaVEAT - an analytic radiative xfer may not be possible without assumptions about emission and absorption mechanisms in the disk. review H- opacity, dust emissivity  mechanisms. what is composition of disk. No conosco.  Another approach is to see if the data are consistent with some simple models. e.g. suppose source I is blackbody. what absorption is consistent with the measured flux of compact source as a function of frequency. keep it simple. e.g. Agnostic approach. Try fitting a 2 layer model with inner, and outer brightness temperature Tin, and Tout. The inner brightness is attenuated. The outer is not. 

 Although the heating of the disk may be complex, the temperature of disk material can be quite well described by a simple radiative transfer model with a hot inner surface of the disk and a cool outer surface, in which the radiation from the inner surface is attenuated by the disk.

%{\color{red} 
We fitted a simple model with a hot inner surface layer of a cavity (excavated by the outflow
and luminosity of SrcI) with brightness temperature $T_{in}$, and an outer surface with brightness temperature $T_{out}$. Radiation from the inner surface is attenuated by the disk. We assume an
opacity which scales as $\nu^{\beta}$:

\begin{multline}\label{radiative-transfer}
T_{obs}(\nu) = T_{in}(\nu) \left( \exp(-\tau_{\nu_0}(\nu/\nu_0))^\beta \right)\\ + T_{out}(\nu)
\end{multline}

%   Tobs(nu) = Tin(nu)( 1 - exp(-tau_nu0(nu/nu0)^beta) + Tout(nu)
% 4 unknowns Tin Tout nu0 and beta
% 5 equations with Tobs at nu = 43 86 99 220 and 340 GHz
% /o/wright/public_html/UG_Research_2020/UG_research
% morph>grep foreach Tmodel_Chi2.csh
%  foreach Tin ( 1000 1250 1500 1750 2000 )
%    foreach Tout ( 400 500 600 )
%      foreach Tau43 ( .1 .25 .5 .75 1 )
%        foreach Beta ( 0 .25 .5 .75 1 1.5 2 )
% # foreach Freq ( 43 86 99 220 340 )

 We searched a parameter space of $T_{in}$, $T_{out}$, $\tau_{43GHz}$, and $\beta$ 
towards the compact source for our observations at 43, 86, 99, 220, and 340~GHz. At 30 mas resolution,
minimum Chisq fits to the observed brightness temperature $T_{obs}$ were in the range, $T_{in}$ = 1000 -1500 K, $T_{out}$ = 400 - 600 K, $\tau_{43GHz}$ = 0.1 - 0.75, and $\beta$ = 0.75 - 2,
with the range of values corresponding to a doubling of Chisq around the minimum. At other positions away from the compact source (which is offset from the center of the disk), the model requires further assumptions about the geometry and heating of the disk. The inner surface may be heated by radiation from SrcI, shocks from the outflow and accreting material, and radiative heating from the outer surface and outflow.
With 30 mas (12 AU) resolution, the observed brightness, $T_{obs}$, averages over several features seen in the SrcI disk, and the interpretation is less clear away from the compact source. Observations with better resolution and sensitivity would enable more detailed modelling.

Our crude two-temperature model is inspired by the similar model used in \citep{Li2017}, and fits the hypothesis that the spectral index is dominated at low frequencies by the H\textsuperscript{-} opacity, and at higher frequencies by the cool dust opacity obscuring the inner disk. 
The implied luminosity
at the inner and outer surfaces with these temperatures are consistent with the luminosity of Src I $\sim$ 10000 L$_{\odot}$.
Modeling the disk as a cylinder with radius 50 AU and length 21 AU (see table 2), the luminosity for a black body with surface temperature in the range 400 - 600 K $\sim$ 2000 - 9500 L$_{\odot}$. For the inner surface, using a radius $\sim$ 40 AU, defined by the AlO emission, with a surface temperature in the range 1000 - 1500 K the luminosity $\sim$ 1600 - 8300 L$_{\odot}$.

 The nearly edge on disk is opaque at IR and sub-millimeter wavelengths (e.g. at 340 and 220 GHz), but at 43 - 90 GHz the internal structure of the disk is revealed. Radiation at IR wavelengths can escape into the wide angle outflow, and may contribute to the excitation of the SiO maser emission \citep{Goddi2009}, and salt emission \citep{Ginsburg2019}.

% Luminosity
% 5.67e-5*3.142*(100*21+2*50**2)*(1.5e13)**2*500**4/3.9e33 =4560
% 5.67e-5*3.142*(100*21+2*50**2)*(1.5e13)**2*400**4/3.9e33 = 1868
% 5.67e-5*3.142*(100*21+2*50**2)*(1.5e13)**2*600**4/3.9e33 = 9457
% 5.67e-5*3.142*(40**2)*(1.5e13)**2*1000**4/3.9e33 = 16444
% 5.67e-5*3.142*(40**2)*(1.5e13)**2*1500**4/3.9e33 = 83251

%{\color{red}

\subsubsection{The Warped Ridge}

The nature of the warped ridge of emission which is observed from 43 to 99 GHz is not clear. The ridge is most prominent at 99 GHz, and manifests only as an extended peak in the brightness distributions at 220 and 340 GHz as shown in figure \ref{fig:B7-B6-B3},  and figure \ref{fig:43-86-99GHz}.
These observations suggest emission from the hot inner parts of the disk, which is absorbed at 220 and 340 GHz by cooler material in the outer disk.
The ridge could represent a warped disk structure, or filaments which are accreting onto  SrcI. 
Such observable asymmetries are expected in the inner regions of circumbinary disks \citep{Avramenko2017}.
Assuming an enclosed mass 15 $M_\odot$,
the period of orbit is 15 years for material at 15 AU radius and 90 years at 50 AU. 
The encounter with BN $\sim$ 500 years ago is sufficient time to wind up remains of filaments which may have been
accreting onto SrcI before the encounter. However, the twisted ridge may trace active accretion in  SrcI.
% using  v [km/s] = sqrt ( G[900 AU/M0] * M[M0]/r[AU] ) 
 % 	calc '2*pi*50*150e6/sqrt(900*15/50)/pi/1e7' = 91.28709 yr.
 %   calc '2*pi*5*150e6/sqrt(900*15/5)/pi/1e7'   = 2.886752 yr
 %  calc '2*pi*15*150e6/sqrt(900*15/15)/pi/1e7' = 15 yr
% OR using Kepler's law
%    calc '15*sqrt((5/15)**3)'  =  2.886752
%    calc '15*sqrt((50/15)**3)' =  91.28709
  
Figure~\ref{fig:86GHz+SiO54} shows SiO $v=0$ $J=5-4$  emission at 217.10498 GHz overlaid on the 86 GHz continuum emission. 
The SiO $J=5-4$ line shows emission at velocities outside the range -10 to +20 km/s seen in the $J=1-0$ and $J=2-1$ lines \citep{Hirota2020}. This high velocity SiO emission extends from close to the mid-plane of the disk, between the cusps of low brightness continuum in the 86~GHz emission at the outer edge of the disk.
The outflow is closest to the center at $\sim$ -20 km/s, and  +30km/s. If we assume that the outflow is attached to the rotation of the disk, then we can estimate the inner radius, $r_{inner} = GM/v^2 \sim$ 20 AU from the footprint of the SiO $J=5-4$ emission.
This radius is consistent with the launching radius for a magneto-centrifugal wind determined from the rotation of the SiO $v=0$ $J=5-4$ outflow, see equation 4 in \citep{Anderson2003},
and consistent with that determined by \citet{Hirota2017}.

%This equation gives the disk rotation rate Q0 in the wind-launching region from measurable quantities at large distances.
%The center velocity of -20 and +30 is 5 km/s and the rotation velocity, v=25 km/s, giving r = 900 x 15 M0 / (25 km/s)$^2$ = 20 AU

% cont_mfs_sl+sl5+sio54_10kms.pdf
% plot cont_mfs_cm03+sio54_10kms.pdf
\begin{figure*}
% trim left bottom right top

\includegraphics[width=2.0\columnwidth, clip, trim=.5cm 4.5cm 0.5cm 1.2cm]{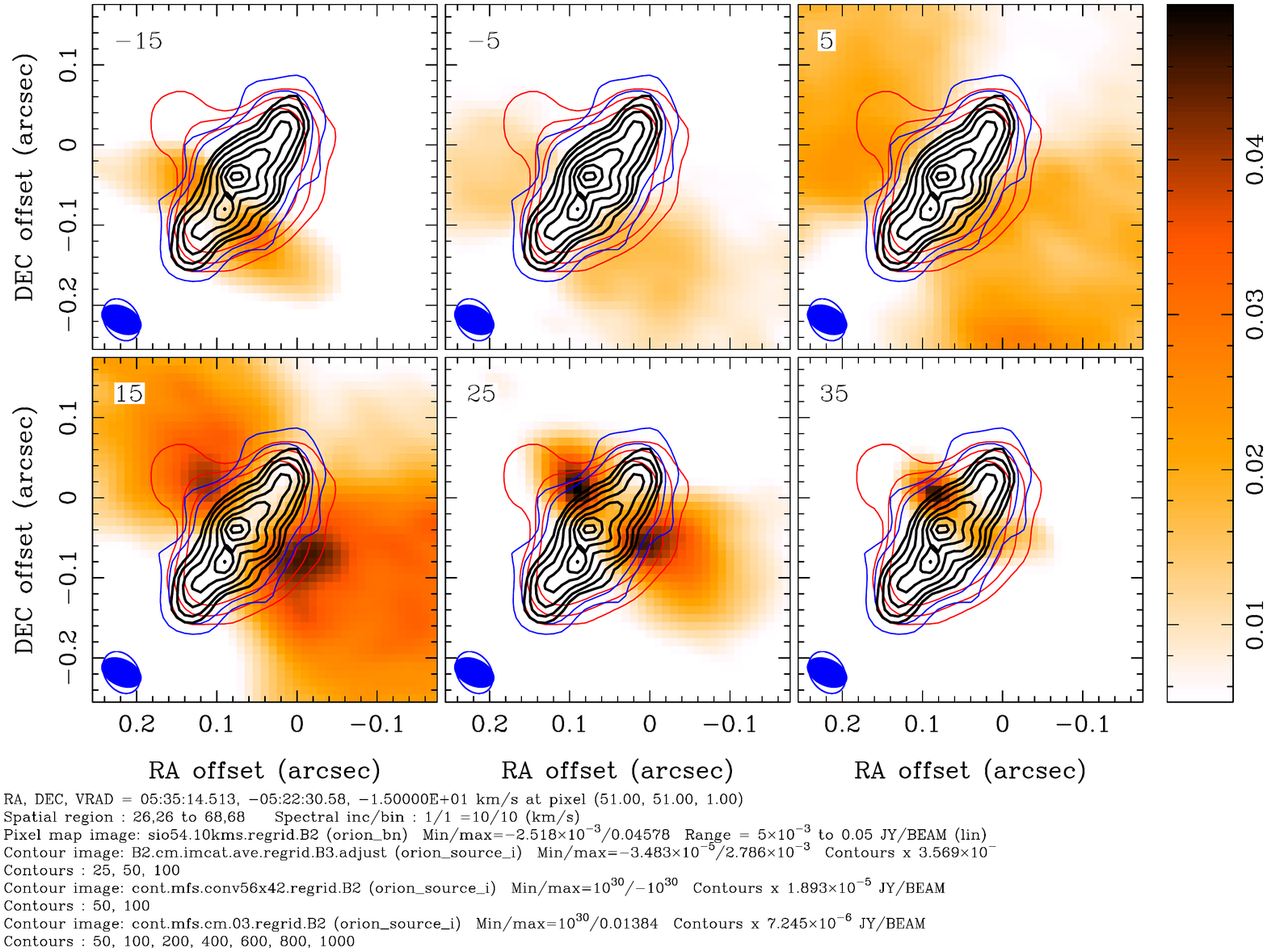}
\caption{SiO $v=0$ $J=5-4$  emission at 217.10498 GHz overlaid on contours of the 99, 86, and 43 GHz continuum emission. The thick black contours at 99 GHz at 50, 100, 200, 300, 400, 500, 600, 700, 800, 900, 1000 K with a 30 mas FWHM show the inner structure in the disk. The blue contours at 86 GHz at 50 and 100K, and red contours at 43 GHz at 25, 50 and 100K are convolved to 56 $\times$ 42 mas FWHM resolution to show the structure at the edge of the disk. The color image shows SiO $v=0$, $J=5-4$ emission at 10 km s$^{-1}$ intervals. Beam FWHM 54 x 34 mas in PA 65 \degr. The beams are shown in the lower left.  
 \label{fig:cont_mfs_cm03+sio54_10kms}}
\end{figure*}

% # /alma_scr/wright/VLA_pol/map2.csh
% # 04jun2021 omit SI and use color for sio. Increase line weight for cont.mfs
% cgdisp in=sio54.10kms.regrid.B2,B2.cm.imcat.ave.regrid.B3.adjust,cont.mfs.conv56x42.regrid.B2,cont.mfs.cm.03.regrid.B2 labtyp=arcsec 'region=arcsec,box(.25,-.25,-.17,.17)' options=full,blacklab,beambl,noepoch,wedge,3val type=p,c,c,c slev=a,3.5688631E-06,a,1.8931789E-05,a,7.2445409E-06 levs3=50,100,200,400,600,800,1000 levs1=25,50,100 levs2=50,100 range=.005,.05,lin,-8 device=B3+B2+conv56x42+sio54.ps/cps cols1=2 cols2=4 cols3=1 lines=2,2,2,4,2,2 csize=1,1,1,1,1
% ps2pdf B3+B2+conv56x42+sio54.ps
% cp B3+B2+conv56x42+sio54.pdf ~/public_html/OriSrcI

In order to investigate the relationship of the SiO emission to
the continuum emission in more detail, we deconvolved the 99 GHz image using
the CLEAN algorithm. The CLEAN models are not unique, but when
convolved by the synthesized beam are consistent with the synthesized image, and a residual image with a Gaussian distribution of pixel values with an RMS noise level 0.4 mJy. Images of the 4 spectral windows which were used to make the 99 GHz continuum image are consistent.
As a further check, we made CLEAN models using an alternative self-calibration based on an SiO v=1 maser feature, instead of the standard calibration.  Both images show the twisted ridge of emission.

Figure~\ref{fig:cont_mfs_cm03+sio54_10kms}  shows SiO $v=0$ $J=5-4$  emission overlaid on the
99, 86 and 43 GHz continuum emission. The 99 GHz emission is convolved to a 30 mas resolution to show the inner structure in the disk. The blue contours at 86 GHz at 50 and 100K, and red contours at 43 GHz at 25, 50 and 100K are convolved to 56 $\times$ 42 mas FWHM resolution to show the structure at the edge of the disk.
The high velocity SiO emission appears quite localized, with a clear footprint in the twisted ridge observed in the continuum emission, whereas the SiO emission between -10 and +20 km/s stands away from the continuum ridge. At the edge of the disk, the high velocity SiO
emission emerges in the cusps of the outer contours at 86 and 43 GHz, suggesting that material is ablated by the outflow.
These observations suggest that the twisted ridge traces the working surface of the disk which is heated by
accretion onto the disk and the disruption of the disk surface by the outflow.

%{\color{orange} 
%The scaling relation }

%\begin{equation}
    %r_{rim} \sim .5 \left( \frac{L_*}{50 L_\odot} \right)^{1/2}
%\end{equation}

%{\color{orange} \noindent was generalized from the \cite{Dullemond-protoplanetary-disks} study on disks, and is applied here to assess the viability of a binary system of proto-stars lurking in the SrcI disk. Using the value of $10^4 L_\odot$ for SrcI, the radius of the rim would be expected at around $r_{rim} \sim$ 7 au \citep{Reid2007} \citep{Plambeck2016}.

%If there is a binary, the rim radius would be about twice the orbital radius \cite{Williams_2011}. This consideration constrains such a radius to less than $\sim$ 3 au. Currently the highest resolution observations are $>$ 12 au in resolution with ALMA. The evidence presented here supports a rim radius $<$ 15 au. Future studies might resolve this inner region and provide conclusive evidence for any binary-related questions.}

\subsection{Spectral Energy Distribution}

There is another potential connection to the twisted ridge not yet mentioned. As highlighted above, the optical depth increases with frequency and the twisted ridge structure appears clearest around $\sim 86\text{-}99$ GHz. 
%The disk appears to have settled into a flared morphology yielding a torus-like shape at larger radii and at the highest frequencies. 
The question remains if the structure evident in the images is directly related to apparent curvature on wide band spectral energy distributions (SEDs) \citep{Plambeck2016}, \citep{Beuther2006}; it is possible that a non-collisional mechanism is exciting the SiO $J=5\text{-}4$ outflow and is related to both the excess flux density at low frequencies, which  \citet{Plambeck2016} attributed to free-free emission (probably H- free-free emission), and the warped ridge. A full analysis of the SED is beyond the scope of this paper.
% Dick's comments:Sec 3.3 -- "The disk appears to have settled into a flared morphology yielding a torus-like shape at larger radii and at the highest frequencies."  When I look at the 220 and 340 GHz images in Fig 2, I see a brick, not a torus.

%Sec 3.3 -- I'm not clear on what is meant by "the observed bump in the SED".  Does this refer to excess flux density at low frequencies, which Plambeck and Wright 2016 attributed to free-free emission (probably H- free-free emission)?  The text suggests that there's a connection between the structure of the disk and the SED, but offers no explanations for that connection.
%}

\subsection{Disk Mass, Accretion, and Outflow}

In this section we discuss accretion and outflow in the disk.
%Although this paper focuses on the structure of the disk,
The circumbinary disk may have an evolving structure with variable accretion and outflows.
The disk mass is estimated to be in the range 0.02-0.2 M$_{\odot}$ \citep{Plambeck2016}. 
%{\color{red} add a few more details of calculation from plambeck2016}
A disk mass of $\sim$ 0.1 M$_{\odot}$ was derived in \citet{Hirota2017}.
Any original disk would have been truncated to the observed radius in the
encounter between Src I and BN  \citep{Moeckel2012}, \citep{Ginsburg2018}, \citep{Bally2020}.
Src I was ejected and is moving through
the debris of the explosion at a velocity $\sim$ 10 km/s.  
Using a velocity $\sim$10 km/s relative to the OMC1 core, the accretion onto the disk from the ambient cloud would be limited to roughly the Bondi-Hoyle rate, which for $n_{H_{2}} = 1 \times 10^7 \; \text{cm}^{-3}$, $\rho \sim 3 \times 10^{-17} \; \text{g} \; \text{cm}^{-3}$, $M = 15 \; M_\odot$, and $v= 10 \; \text{km} \; \text{s}^{-1}$.
%relative to the interstellar background is given by

\begin{equation}
    \dot{M} = \lambda_* 4 \pi \rho \frac{\left(GM\right)^2}{\left(v^2 + c_s^2\right)^{3/2}} \approx 3 \times 10^{-5} M_\odot \; \text{yr}^{-1}
\end{equation}

\noindent where $c_s = \sqrt{ k_B T / \mu m_p }$ is the ambient thermal (sound) speed, which is in the range $1\text{-}5 \; \text{km} \; \text{s}^{-1}$  for $T=10\text{-}1000 \; \text{K}$, and $\mu = 2 - 3$ with $\lambda_*$ being a dimensionless constant of order unity \cite{shu1991physics}.
% mchw - changed \mu = 1/2 to 2-3 for molecular H2 + some He

%$\mathrm{n}(\mathrm{H}_{2})=1.0\mathrm{e}{7} \ \mathrm{cm}^{-3}, \ \rho \sim 4.5 \mathrm{e}{-17} \ \mathrm{g}/\mathrm{cm}^{3}, \  \mathrm{M}=15 \ \mathrm{M}_{\odot}, \ \mathrm{and} \ \mathrm{v}=10 \ \mathrm{km}/\mathrm{s}$ is
%$\mathrm{dM}/\mathrm{dt} \sim \pi \mathrm{G}^2 \mathrm{M}^2 \rho / \mathrm{V}^3 = 8.5\mathrm{e}{-6} (\mathrm{M_{\odot}}/\mathrm{year})$.

Considering that SrcI does not appear to be in the densest part of OMC1, this may be an upper bound.
% If material from the BN/SrcI encounter is co-moving with SrcI, then the accretion rate could be higher. 
Accretion from the filamentary debris of the BN/SrcI explosion could produce large variations in the accretion rate. 
SrcI may pass through multiple filaments of explosion debris at different relative velocities. For co-moving material the B-H accretion rate could be as high as $3 \times 10^{-3} M_\odot \; \text{yr}^{-1}$, assuming a sound speed $\sim$ 2 km/s. Accretion variations are to be expected as SrcI passes through the filamentary debris of the BN/SrcI explosion. The crossing time for a 1'' ( 400 AU) filament at 10 km/s $\sim$ 200 yr.

% B-H accretion rates for 15 Mo and 10, 2 and 3 km/s
%>>> 4*3.14*3e-17*(6.67e-8*15*2e33)**2/10e5**3*3.15e7/2e33
% = 2.3762144334599997e-05
%>>> 4*3.14*3e-17*(6.67e-8*15*2e33)**2/2e5**3*3.15e7/2e33
% = 0.0029702680418250004
%>>> 4*3.14*3e-17*(6.67e-8*15*2e33)**2/3e5**3*3.15e7/2e33
% = 0.0008800794198

 \citet{Wright1995} estimated the mass of the outflow from images of the SiO J=2-1 , v=0 line.
 %{\color{red} see \citet{Wright1995} sections 4.1 and 4.3}
 Using a solar abundance of Si, n(SiO)/n(H$_2$) $= 6\times 10^{-5}$,
 the total mass in the outflow was estimated to be $\sim 5 \times 10^{-3}$ M$_{\odot}$. Using an outflow velocity 18 km/s, the mass loss rate $\sim 10^{-5}$ $\mathrm{M_{\odot}}/\mathrm{yr}$ and the kinetic luminosity $\sim$ 1 L$_{\odot}$.
 % >>> 0.5*5e-5*2e33/3.16e7*10e5**2/3.9e33 = 0.40 L0  ; but we use 18 km/s for consistency
 Since the Si abundance and degree of masing in the SiO J=2-1 , v=0 line is unknown, this is only a rough estimate.
 \citet{Greenhill2013}, estimate a loss rate $\sim 5 \times 10^{-6}$ $\mathrm{M_{\odot}}/\mathrm{yr}$ from the SiO J=1-0, v=0 outflow.

 \citet{Lopez-Vazquez2020}  derive an outflow mass from the $^{29}$SiO J=8-7, v=0 line. Using an SiO abundance $1.2 - 2.4 \times 10^{-7}$ relative to H$_2$ \citep{Ziurys1987}, they estimate the mass of the molecular outflow 0.66 - 1.3 M$_{\odot}$, and a kinematic age 130 yr. This implies a mass loss rate in the outflow $\sim 5 -10 \times 10^{-3}$ M$_{\odot}$/yr.
 \citet{Lopez-Vazquez2020} note that this estimate leads to some problems:
 %which can be avoided by using a high outflow-to-accretion ratio:
 
 %However what is the KE of the higher mass outflow ?
% >>> 0.5*5e-3*2e33/3.16e7*10e5**2/3.9e33 = 40.57 L0  -- still is OK

i) The outflow depletes disk mass unless the accretion rate is high, and

ii) The accretion luminosity exceeds $10^4$ L$_{\odot}$ if the
accreting material falls onto the stars.

  \citep{Ziurys1987} determined the SiO abundance 1.2 - 2.4 $\times 10^{-7}$ relative to $H_2$ from a N($H_2$) column density $1-2 \times 10^{24} cm^{-2}$, and N(SiO) $2\times 10^{17}$ cm$^{-2}$ for a 10'' region
 centered on SrcI. However, the explosion 550 years ago could have vaporized all dust.

 SiO is abundant in the debris from the explosion. (see figure 17 in \citep{Wright2020}).  
The Solar abundance of Si is $\sim 3 \times 10^{-5}$ relative to H.  It is possible that most Si ended up in SiO. If so, the $H_2$ mass in the SrcI outflow could be much lower than \citet{Lopez-Vazquez2020} estimate.    
This would lower the mass-loss rate estimate correspondingly.  But, if the mass estimate ($\sim$ 1 M$_{\odot}$) and time-scale ($\sim$ 130 year) stand, then there must be a better explanation.   

Episodic accretion is a possibility.  The SiO $v=0$ $J=5-4$ images in figure~\ref{fig:86GHz+SiO54} suggest
 different outflow events for the bright compact, high velocity emission, and the more extended emission
 between -10 to +20 km/s. 
 % 05may convolved J=2-1 by VLA beam reduces extent of the J=2-1 outflow due to lack of short spacings. Therefore remove the following text: The SiO v=0 J=2-1 outflow extends $\sim$ 1000 AU from SrcI whereas the J=1-0 outflow $\sim$ 500 AU, although both have similar velocity widths \citep{Hirota2020}. Using a 18 km/s outflow velocity, the dynamic timescales for the J=2-1 and J=1-0 outflows are 260 and 130 yr. respectively.
% >>> 480*150E6/18/3.16e7 =126.58
%>>> 1000*150E6/18/3.16e7 =263.7
%>>> 500*150E6/18/3.16e7 = 131.85
% Although it is possible to seek other explanations for 
 The different extents of the SiO $v=0$ $J=5-4$  emission, could
 have resulted from different outflow events.  Shocks from multiple outflow events may also be responsible for
 the abrupt change in the SiO $v=0$ $J=2-1$  emission outflows from a rotating column into a wide angle outflow $\sim$ 50 AU from the SrcI disk (see figures 1 and 2 in \citet{Hirota2020}).
 Another possible interpretation of the collimated and wide angle outflows are different driving mechanisms at different launching regions \citep{Machida2014}
%, ApJL, 796, L17, 
\citep{Matsushita2017}
%, MNRAS 470, 1026).
 
There is direct evidence from IR photometry 
%{\color{red} needs reference}
and the morphology of the outflows that mass-loss rate, and therefore accretion, is highly variable. \citet{Bally2020} argued that these parameters likely follow a ``1/f" distribution - small variations on short time scales and very large variations on long-time scales.  

The  MJy $H_2O$ maser flare in Orion \citep{Matveenko1982} appears to mark a spot in the SW portion of the SrcI outflow.  It may trace a shock where fast ejecta ran into slower moving material.  Thus, it might be an indirect indicator of mass-loss-rate and/or velocity variability.

If the inflow-rate in the disk is 0.01 M$_{\odot}$/yr, why is the luminosity only L$\sim$10$^4$ L$_{\odot}$?

Perhaps the accretion merely transfers gas from a larger radius to a smaller one without landing on the central star or binary.  The inflow might be prevented from falling onto the central object, perhaps by magnetic stress or a centrifugal barrier if the disk wind is not sufficiently efficient in removing angular momentum.

Alternatively, if the central object is accreting at a very high rate ($> 1\times 10^{-3} M_{\odot}$/year), a 15 M$_{\odot}$ protostar will have an AU-scale photosphere (e.g. Hosokawa \& Omukai 2009).   Accretion at rate of 0.01 M$_{\odot}$/year onto a 15 M$_{\odot}$ object with an R $\sim$ 1 AU photosphere would have a luminosity of $\sim 2\times 10^4 L_{\odot}$. This is NOT inconsistent with the Teff $\sim$ 4,000 K photospheric temperature found by \citet{Testi2010} from the Src I IR reflection nebulosity.

If the central object is a merger remnant, left over from the dynamical interaction ~550 years ago, its photosphere may be especially bloated because it takes a K-H time scale time to radiate away the energy liberated by a collision between two stars.

\subsection{Evidence for a binary in SrcI}
The warped inner disk structure may be direct evidence that SrcI is a binary.
A possible explanation for our observations is an unequal binary that is experiencing episodic accretion. \citet{Dunhill2015} describe numerical simulations of the circum-binary disc
around the young, eccentric stellar binary HD 104237. They find that the binary clears out a
large cavity in the disc, driving a significant eccentricity at the cavity edge. 
The disk precesses around the binary, which for HD 104237 corresponds to a
precession period of 40 years.  The accretion from the disk onto the binary changes with this precession, resulting in a periodic accretion variability determined by the parameters of the binary and its orbit. They find an order of magnitude changes in the accretion rate onto the binary.

 \citet{Dunhill2015} give the disk precession period as 360/0.48 $\times T_{binary}$ = 40 years for HD 104237. T$_{\mathrm{binary}}=\sqrt{\frac{\mathrm{R}^3}{\mathrm{M}}}$ where R is in AU and M is in M$_{\odot}$.  The favored scenario for the formation of the binary in the BN/SrcI explosion  gives an orbital separation 1–9 AU, with an average of 4 AU in the
 last 50 years of the simulation \citep{Goddi2011b}. For R = 1 AU, 2 AU, and 4 AU, and
using a mass of 15 M$_{\odot}$, the precession periods are  $\sim$ 194 yr, $\sim$ 578 yr, and $\sim$ 1549 yr respectively. Although we don't see any evidence of precession in the SiO outflow in the last 500 years, the faster precession would be relevant to the idea of episodic accretion.
 From extensive N-body simulations 
 \citet{Goddi2011b} and \citet{Moeckel2012}
 concluded that the dynamical history of BN/KL cannot be reconstructed with unequal-mass binaries.
% Hirota-san email 26Apr2021
% Goddi email 11May2021
\citet{Goddi2011b} considered N-body simulations with up to 11 stars and masses in the range 0.5-10 M$_{\odot}$. In order to have a 10 M$_{\odot}$ star ejected (BN), the binary must have been comprised of the largest stars in the sample (10+10 M$_{\odot}$). 
\citet{Moeckel2012} investigated the disk survival in the presence of a stellar encounter between a binary and intruder.
%, we considered an equal-mass binary 10+10 Ms and an intruder (BN) of 10 Msun, and an unequal-mass binary 10+3 Ms and an intruder of 10 Msun. 
% We ran over 10^4 simulations. 
In the case of an unequal-mass binary 10+3 M$_{\odot}$, stellar ejections were possible only by swapping the 3 M$_{\odot}$ binary member with the 10 M$_{\odot}$ intruder, resulting in a 10+10 M$_{\odot}$ binary, but there was no case with a 10+3 M$_{\odot}$ binary and 10 M$_{\odot}$ runway. 
However, an accretion event, after the compact binary ejection, which initiates an SiO outflow is not excluded from these simulations. \citet{Farias2018} made $\sim 10^7$ N-body simulations including binary-binary interactions. They conclude that the BN/SrcI requires a SrcI mass $\sim 20 M_{\odot}$, but note that other possible initial combinations remain to be explored.
% consistent with the simulations ? A bit like an FU Orionis event - which I also thought about referring to.

%Yes this is not excluded from the simulations. But we did not go that far as to investigate the consequences of the encounter in terms of triggering accretion. But this scenario is considered in Goddi, C., Ginsburg, A., et al.\ 2020, where the different SiO outflow axes on different scales (i.e. time) are likely a consequence of episodic accretion, like an FU Orionis event, but for higher masses.
\citet{Rasusa2020} studied the evolution of co-planar binaries and the circum-binary discs, using 3D Smoothed Particle Hydrodynamics simulations.
 They find that, even with circular initial conditions, all discs with mass ratios q $>$ 0.05 develop eccentricity, and that the disc eccentricity grows abruptly after $\sim$ 400-700 binary orbits, and is associated with a very small increase in the binary eccentricity.

\citet{Goddi2020} present evidence for episodic accretion from the structure of the SiO jet in W51-North, where different SiO outflow axes on different scales are likely a consequence of episodic accretion, like an FU Orionis event, but for higher masses.
It would be interesting if we see possible evidence of episodic events 
in Source I. As suggested by \citet{Hirota2014}, %, PASJ, 66, 106
the super maser flare could have a periodicity of 13 years (1985, 1998, 
2011), although there was no confirmed flare event in 1973. In order to 
confirm the periodicity, we are waiting for the next flare. Even if  there is no regular periodicity, it could suggest time-variable star-formation activities either in Source I or unknown YSO in the Compact Ridge \citep{Hirota2014}.

 Based on these models, the compact source that we observe at 43 GHz could be the current site of accretion from the disk onto the binary, and driving the most recent outflows. An unequal binary could provide episodic accretion, as well as provide a possible explanation for the warped ridge of emission.
 
  The thick green contours at SI=1.7, 2 and 2.3 in figure~\ref{fig:86Hz-43GHz_SI} delineate the region of lower SI along the minor axis
of the disk. The twist in the SI contours is in the direction of disk rotation suggesting that material ablated 
from the disk is caught up in the disk rotation. The twist has a footprint close to the compact source in the warped ridge, and extends to the edge of the disk with a twist $<$ 1 turn.
The 43 GHz and 86 GHz observations were obtained  within a $\sim$ 5 month period (see Table 1), 
so the offset in the images from the proper motion of SrcI is less than $\sim$ 1 au.
 For an outflow velocity $\sim$ 18 km/s, the time for material to reach the edge of the disk $\sim$ 5 yr
( 18 au and 18 km/s are both measured along the line of sight).
% 18 au x 150e6/18km/s/3e7s  = 5 yr
Assuming an enclosed mass 15 M$_{\odot}$, the period of orbit is 15 years for material at 15 AU radius, and only 3 yr at 5 AU.
In this case, we might expect to see changes in the structure of the disk within a few years. Future observations could reveal changes in the structure of the outflows or disk in SrcI. A more accurate determination of accretion events would also allow us to set constraints on the binary parameters.
 A warped disk is expected in the presence of a central binary, and this could be direct evidence that SrcI is a binary.
  The formation of tight binaries through decay of non-hierarchical multiple systems is a clear mechanism for evolution from wide-separation to close-separation binaries within star-forming regions, which may help solve the problem of a lack of short-period binary systems in early-stage high-mass star-forming regions \citep{Sana2017}.
  %(https://ui.adsabs.harvard.edu/abs/2017A%26A...599L...9S/abstract).

% --- table of observations --- %
\begin{deluxetable*}{cccccc}
\tabletypesize{\small}
\tablecaption{Observations}
\tablecolumns{6}
\tablenum{1}
\tablehead{
\colhead{freq} &
\colhead{project code} &
\colhead{date} &
\colhead{time} &
\colhead{synth beam}  &
\colhead{baseline} \\
\colhead{(GHz)} & & & 
\colhead{(min)} &
\colhead{(milliarcsec)} &
\colhead{(meters)}
}
\startdata
 43  &  VLA/18A-136     &  2018-03-06  &  291  &   39$\times$34 at PA 1  & 500 - 36600  \\
 86 and 99  &  2017.1.00497.S	&  2017-10-12  &  158  &   45$\times$36 at PA 47  & 40 - 16200 \\
 %216-220  & 2013.1.005446.S  & 2014-12 to 2015-04  &   15    &   1500$\times$930 at PA -8 & 14 - 330 \\
224  &  2016.1.00165.S	&  2017-09-19  &   44  &   39$\times$19 at PA 66  & 40 - 10500 \\
340  &	2016.1.00165.S  &  2017-11-08  &   45  &   26$\times$11 at PA 58  & 90 - 12900 \\
%350  &	2012.1.00123.S  &  2014-07-26  &   24  &   276$\times$260 at PA 85 & 30 - 730 \\
\enddata
\end{deluxetable*}

% -------------------------------------- TABLE -----------------------------------------------
\begin{deluxetable*}{CCCC}   % capitals indicate math mode
\tabletypesize{\small}
\tablecaption{Measured sizes and flux densities for source I}
\tablecolumns{4}
\tablenum{2}
\tablehead{
\colhead{freq} &
%\colhead{beam} &
\colhead{deconvolved size} &
\colhead{integrated flux} \\
\colhead{(GHz)} &
%\colhead{(arcsec, PA)} &
\colhead{(arcsec, PA)} &
\colhead{(mJy)}
}
\startdata
%
% fit to VLA_pol/B2.cm.imcat.ave
43 &
%0.06 \times 0.04, 33\degr &
0.099 \pm 0.002 \times 0.057 \pm 0.002, -40.4\degr \pm 2.7\degr &
10 \pm 1 \\
%
% fit to OriPol2018/86GHz.mfs.I.cm
86 &
%0.04 \times 0.04, 82\degr &
0.142 \pm 0.005 \times 0.044 \pm 0.001, -38.0\degr \pm 1.0\degr &
48 \pm 5 \\
%
% fit to OriPol2018/cont/cont.mfs.cm
99 &
%0.04 \times 0.04, 47\degr &
0.151 \pm 0.005 \times 0.044 \pm 0.002, -37.8\degr \pm 1.3\degr &
58 \pm 6 \\
%
% fit to B6.ms0.mfs.cm
224 &
%0.04 \times 0.02, 66\degr &
0.197 \pm 0.003 \times 0.042 \pm 0.003, -37.3\degr \pm 0.4\degr &
256 \pm 25\\
%
% fit to X438c.cont.cm
340 &
%0.03 \times 0.01, 58\degr &
0.234 \pm 0.005 \times 0.042 \pm 0.002, -37.4\degr \pm 0.3\degr &
630 \pm 63 \\
\enddata\label{table2}
\end{deluxetable*}
% --- table of 86 GHz continuum frequencies --- %
\begin{deluxetable*}{cccc}
\tabletypesize{\small}
\tablecaption{ 86 GHz continuum frequencies}
\tablecolumns{4}
\tablenum{3}
\tablehead{
\colhead{spectral window} &
\colhead{frequency (GHz) } &
\colhead{frequency (GHz) } &
\colhead{bandwidth (MHz) } \\
}
\startdata
0  & 86.263704 &  86.223543  & 18.3 \\
1  & 86.867220 &  86.827058  & 18.3 \\
8  & 86.263376 &  86.223215  & 18.3 \\
9  & 86.866891 &  86.826730  & 18.3 \\
2  & 84.763768 &  84.728856  & 23.4 \\
4  & 85.776890 &  85.741978  & 23.4 \\
5  & 85.184697 &  85.149785  & 23.4 \\
10 & 84.763444 &  84.728532  & 23.4 \\
12 & 85.776567 &  85.741655  & 23.4 \\
\enddata
\end{deluxetable*}

\section{SUMMARY}

 %{\color{red} %mchw. suggest MAKE AN ITEMIZED LIST OF NEW RESULTS IN THIS PAPER

1. We present a new image of the source I disk at 86 GHz with $\sim$ 30 mas resolution.

2. Images at 43, 86 99, 220, and 340 GHz reveal significant structure in the source I disk:
 a compact source most prominent at 43 GHz, and a warped ridge of emission at 86 and 99 GHz.
 The warped inner disk structure may be evidence for a non axis-symmetric potential within the disk, or
 recent accretion events.
 % that SrcI is a binary.

3. The spectral index, SI $\sim$ 3 at the ends of the disk along the major axis, 
 and there is an extended region with SI $<$ 2 along minor axis, and along the major axis ridge. 

4. The observed spectral index of the disk can be understood as a hot inner core and cool, dusty outer envelope first discussed in {\citep{Ginsburg2018}}.

5. We fitted a simple model with hot inner structure heated by source I, with brightness temperature $T_{in}$, and an cooler outer surface with brightness temperature $T_{out}$. Radiation from the inner structure is attenuated by the outer disk. Assuming an opacity which scales as $\nu^{\beta}$, the best fits to the observed brightness temperature
towards the compact source for our observations at 43, 86, 99, 220, and 340~GHz at 30 mas resolution
are in the range, $T_{in}$ = 1000 -1500 K, $T_{out}$ = 400 - 600 K, $\tau_{43GHz}$ = 0.1 - 0.75, and $\beta$ = 0.75 - 2.
The warped ridge is consistent with emission from the hot inner parts of the disk, which is absorbed at 220 and 340 GHz by cooler material in the outer disk.

6. High velocity SiO $v=0$ $J=5-4$  emission has a footprint in the warped ridge.
We suggest that the ridge traces the working surface on the disk which is heated by
accretion onto the binary and the disruption of the inner disk surface by the outflow.
 
7. The disk structure may be evolving,
with multiple accretion and outflow events. We identify two sources of variability; variable accretion onto the circum-binary disk as source I travels through the filamentary debris from the BN-Source I encounter $\sim$ 550 yr ago, and episodic accretion from the circum-binary disk onto the binary which may trigger multiple outflows.
%% If you wish to include an acknowledgments section in your paper,
%% separate it off from the body of the text using the \acknowledgments
%% command.

\acknowledgments
We thank Gibor Basri, Richard Plambeck and Goran Sandell for insightful discussions which have improved this paper.
 We thank the referee for a careful reading which has improved the presentation of this paper.
This paper makes use of ALMA data listed in table 1.
%ADS/JAO.ALMA\#2012.1.00123.S, ADS/JAO.ALMA\#2013.1.00546.S,
ADS/JAO.ALMA\#2016.1.00165.S,
ADS/JAO.ALMA\#2017.1.00497.S. 
 ALMA is a partnership of ESO (representing its member states), NSF (USA) and NINS (Japan), together with NRC (Canada), MOST and ASIAA (Taiwan), and KASI (Republic of Korea), in cooperation with the Republic of Chile. The Joint ALMA Observatory is operated by ESO, AUI/NRAO and NAOJ. 

The National Radio Astronomy Observatory is a facility of the National Science Foundation operated under cooperative agreement by Associated Universities, Inc.
TH is financially supported by the MEXT/JSPS KAKENHI Grant Numbers 
17K05398, 18H05222, and 20H05845.

%“The National Radio Astronomy Observatory is a facility of the National Science Foundation operated under cooperative agreement by Associated Universities, Inc."
%3.1.2 ALMA Acknowledgement
%Additionally, publications making use of ALMA data must include the following statement in the acknowledgement:
%"This paper makes use of the following ALMA data: ADS/JAO.ALMA#2011.0.01234.S. ALMA is a partnership of ESO (representing its member states), NSF (USA) and NINS (Japan), together with NRC (Canada), MOST and ASIAA (Taiwan), and KASI (Republic of Korea), in cooperation with the Republic of Chile. The Joint ALMA Observatory is operated by ESO, AUI/NRAO and NAOJ. "

%% To help institutions obtain information on the effectiveness of their 
%% telescopes the AAS Journals has created a group of keywords for telescope 
%% facilities.
%
%% Following the acknowledgments section, use the following syntax and the
%% \facility{} or \facilities{} macros to list the keywords of facilities used 
%% in the research for the paper.  Each keyword is check against the master 
%% list during copy editing.  Individual instruments can be provided in 
%% parentheses, after the keyword, but they are not verified.

\vspace{5mm}
\facilities{ALMA, VLA}

%% Similar to \facility{}, there is the optional \software command to allow 
%% authors a place to specify which programs were used during the creation of 
%% the manusscript. Authors should list each code and include either a
%% citation or url to the code inside ()s when available.

\software{Miriad \citep{Sault1995}}

\bibliographystyle{aasjournal}
\bibliography{SrcI.bib}

%% Include this line if you are using the \added, \replaced, \deleted
%% commands to see a summary list of all changes at the end of the article.
%\listofchanges

\end{document}